\documentclass[global,twocolumn,final,a4paper]{svjour}
\usepackage{times}
\journalname{Applied Physics B}

\usepackage{amssymb,amsmath,amsfonts,amsbsy,bbm,enumerate}

\providecommand{\bra}[1]{{\langle{#1}\rvert}}
\providecommand{\ket}[1]{{\lvert{#1}\rangle}}
\providecommand{\bracket}[2]{{\langle{#1}|{#2}\rangle}}

\providecommand{\abs}[1]{{\lvert{#1}\rvert}}
\newcommand{\ketbra}[2]{{\ket{#1}\!\bra{#2}}}
\newcommand{\proj}[1]{{\ketbra{#1}{#1}}}
\newcommand{\id}{{\mathbbm 1}}
\newcommand{\purestate}[1]{{\Pi_\ket{#1}}}

\newcommand{\Hilbert}{{\mathcal H}}
\newcommand{\LinearOpSpace}[1]{{\mathcal B({#1})}}
\newcommand{\DensityOpSet}[1]{{\mathcal B_+({#1})}}

\newcommand{\Min}{\mathrm{in}}
\newcommand{\Mout}{\mathrm{out}}
\newcommand{\Haux}{{\Hilbert_\mathrm{aux}}}
\newcommand{\traux}{{\tr_\mathrm{aux}}}

\newcommand{\M}{{\mathcal M}}
\newcommand{\LinearOp}{{\chi}}

\newcommand{\td}[2]{{d({#1},{#2})}}
\newcommand{\wcd}[2]{{\mathcal D({#1},{#2})}}

\DeclareMathOperator{\tr}{tr}
\DeclareMathOperator{\Span}{span}
\DeclareMathOperator{\rank}{rank}
\DeclareMathOperator{\supp}{supp}
\DeclareMathOperator{\Pur}{pur}
\DeclareMathOperator{\Id}{ID}

\begin{document}
\title{Purifying and Reversible Physical Processes}
\author{Matthias Kleinmann \and Hermann Kampermann \and Tim Meyer,
        and Dagmar Bru\ss}
\institute{Institut f\"ur Theoretische Physik,
           Heinrich-Heine-Universit\"at D\"usseldorf,
           D-40225 D\"usseldorf,
           Germany}
\date{\today}
\maketitle

\begin{abstract}
Starting from the observation that reversible processes cannot increase the 
 purity of any input state, we study deterministic physical processes, which 
 map a set of states to a set of pure states.
Such a process must map any state to the {\em same} pure output, if purity is 
 demanded for the input set of {\em all} states.
But otherwise, when the input set is restricted, it is possible to find 
 non-trivial purifying processes.
For the most restricted case of only two input states, we completely 
 characterize the output of any such map.
We furthermore consider maps, which combine the property of purity and 
 reversibility on a set of states, and we derive necessary and sufficient 
 conditions on sets, which permit such processes.
\end{abstract}

\section{Introduction}
The notion of a pure quantum state plays a special role in quantum information 
 theory.
Many problems -- such as separability or the existence of a particular quantum 
 protocol -- can easily be solved, if one restricts the problem to pure quantum 
 states only.
On the other hand mixed states endow quantum systems with many properties (such 
 as bound entanglement), that cannot be found for systems described by pure 
 states.
In our contribution we investigate physical processes which transform a given 
 set of mixed states to a set of pure states.
If such a process exists, then it may e.g. be possible to infer from the 
 properties of the pure output states some properties of the input states.
Such a kind of conclusion is particular powerful, if the purifying map can be 
 chosen to be reversible, since then one can consider the set of pure output 
 states and the set of input states as physically equivalent.

We consider {\em deterministic} physical processes, i.e., processes which map 
 any ~ possible ~ input ~ quantum ~ state ~ with probability one to a 
 corresponding output quantum state.
The states of the input quantum system are represented by density operators 
 $\rho_\Min\in \DensityOpSet{\Hilbert_\Min}$, i.e., positive semidefinite 
 operators with trace one acting on the finite-dimensional complex vector space 
 $\Hilbert_\Min$.
Analogously $\rho_\Mout\in \DensityOpSet{\Hilbert_\Mout}$ shall represent the 
 set of states of the output quantum system.
Any deterministic physical process can be written as a completely positive and 
 trace preserving linear (CPTP) map $\Lambda\colon 
 \LinearOpSpace{\Hilbert_\Min}\rightarrow \LinearOpSpace{\Hilbert_\Mout}$, 
 where $\LinearOpSpace{\Hilbert_{\Min, \Mout}}$ denotes the space of linear 
 operators on $\Hilbert_{\Min, \Mout}$.
In this language, the trace preserving condition reflects the fact that we 
 restrict our considerations to deterministic processes.
In Section~\ref{s18042} we will argue that this restriction is indeed necessary 
 to have a proper definition of a purifying map.

A reversible process is a physical process, where the action of the process on 
 any physical state can be undone by another physical process, i.e., a CPTP map 
 $\Lambda$ is reversible if one can find an inverse map $\Lambda'$ which is 
 also CPTP and satisfies $(\Lambda'\circ \Lambda)[\rho_\Min]= \rho_\Min$ for 
 any density operator $\rho_\Min$.
The most common example are unitary processes $E_U\colon \LinearOp\mapsto 
 U\LinearOp U^\dag$, where $U$ is a unitary transformation, $UU^\dag= \id$.
Here obviously the inverse map is given by $(E_U)'= E_{U^\dag}$.
Another class of reversible processes that is important for our purposes is 
 described by
\begin{equation}\label{e18547}
 E_\sigma\colon
   \LinearOpSpace{\Hilbert_\Min}\rightarrow
     \LinearOpSpace{\Hilbert_\Min\otimes \Haux}\colon
 \LinearOp\mapsto \LinearOp\otimes \sigma,
\end{equation}
 where $\sigma\in \DensityOpSet\Haux$ is some arbitrary density operator.
The inverse map for this process is the partial trace over the auxiliary 
 system, $(E_\sigma)'= \traux$.
A remarkable property of this inverse map is, that it does not depend on 
 $\sigma$ and hence cannot be reversible.
Note that neither the action of $E_U$ nor the one of $E_\sigma$ increases the 
 purity $\tr(\rho^2)$ of any density operator $\rho$.
Indeed a process, which is reversible on the set of all states cannot increase 
 the purity of even a single state:
Let us first consider reversible maps, for which the reverse map is the partial 
 trace (e.g. $E_\sigma$).
For such a reversible map $\Lambda$, the output of any pure state 
 $\purestate\phi\equiv \proj\phi$ must be $\Lambda[\purestate\phi]= 
 \purestate\phi\otimes \sigma_\phi$ for some state $\sigma_\phi$.
For any state $\rho_\Min$ we find with the spectral decomposition $\rho_\Min= 
 \sum_i p_i\purestate{\lambda_i}$ that due to linearity, 
 $\tr(\Lambda[\rho_\Min]^2)= \sum p_i^2\tr(\sigma_{\lambda_i}^2)\le 
 \tr(\rho_\Min^2)$, i.e., no state can become purer by the action of $\Lambda$.
Now consider a general reversible map $\Lambda$.
For the reverse process $\Lambda'$, by virtue of Stinespring's dilation theorem 
 \cite{Stinespring:1955PAMS,Nielsen:2000} one can write the most general form 
 of a CPTP map,
\begin{gather}
 \Lambda'\colon \LinearOp\mapsto \traux U(\LinearOp\otimes 
  \purestate{\mathrm{anc}})U^\dag.
 \intertext{From this we define}
\label{e25670}\begin{split}
 \Gamma_{\Lambda',\Lambda;\ket{\mathrm{anc}}}\colon&
  \LinearOpSpace{\Hilbert_\Min}\rightarrow \LinearOpSpace{\Hilbert_\Min\otimes
    \Haux}
    \\\colon&
  \LinearOp\mapsto U(\Lambda[\LinearOp]\otimes \purestate{\mathrm{anc}})U^\dag.
\end{split}\end{gather}
The inverse map of $\Gamma_{\Lambda',\Lambda}$ obviously is 
 $(\Gamma_{\Lambda',\Lambda})'= \traux$ and by construction, 
 $\tr(\Gamma_{\Lambda', \Lambda}[\rho_\Min]^2)= \tr(\Lambda[\rho_\Min]^2)$ 
 holds for all $\rho_\Min$.
Using the previous result, we find 
 $\tr(\Gamma_{\Lambda',\Lambda}[\rho_\Min]^2)= \tr(\Lambda[\rho_\Min]^2)\le 
 \tr(\rho_\Min^2)$, i.e., no state can become purer by the action of a 
 reversible process.

Since a process that is reversible on all states cannot improve the purity of 
 any state, one would guess that, on the other hand, a process which maps all 
 states to a pure state cannot be reversible for any state.
Such a process is called a purifying process, i.e., ~ a ~ CPTP ~ map ~ 
 $\Lambda$ ~ is ~ purifying, ~ if ~ $\tr(\Lambda[\rho_\Min]^2)= 1$ holds for 
 any input state $\rho_\Min$.
The action of such a map, indeed, has to map any state to the same pure output 
 state:
Suppose $\Lambda[\rho_1]\ne \Lambda[\rho_2]$.
Then for $\rho_3= (\rho_1+ \rho_2)/2$ we find $\Lambda[\rho_3]= 
 (\Lambda[\rho_1]+ \Lambda[\rho_2])/2$, which only can be pure, if 
 $\Lambda[\rho_1]= \Lambda[\rho_2]$ in contradiction to our assumption.
Thus a purifying process must destroy any information of the input state and 
 cannot be reversible at all.

So, the properties of reversibility and output purity are completely 
 incompatible for a physical process, if one demands these properties to hold 
 on all possible input states.
Our approach now is to require these properties only on a certain subset of 
 states $\M\subset \DensityOpSet{\Hilbert_\Min}$.
In Section~\ref{s14898} we analyze the properties of maps, which map at least 
 two mixed states to pure states.
The result of this investigation will completely characterize any such map.
As an application of this result we will provide a lower bound on the trace 
 distance of any two product states $\rho_1\otimes \sigma_1$ and $\rho_2\otimes 
 \sigma_2$.
In a brief excursion in Section~\ref{s18042} we will show that, if we allow 
 probabilistic processes, non-trivial examples of reversible and purifying 
 processes can easily be constructed.
But we will also show that the definition of a probabilistic process to some 
 extent contradicts the properties of a purifying process.
In Section~\ref{s29562} we will then characterize any set of states, for which 
 a deterministic reversible and purifying map exists and discuss in some detail 
 the structure of such sets.
Finally, we conclude in Section~\ref{s10522}.

\section{Purifying Processes of two states}\label{s14898}
In the previous analysis we ruled out the possibility of a non-trivial process, 
 which takes all states $\DensityOpSet{\Hilbert_\Min}$ to a corresponding pure 
 state in $\DensityOpSet{\Hilbert_\Mout}$.
So the question arises, to what extent this also holds if one demands a pure 
 output only for a subset of states $\M\subset \DensityOpSet{\Hilbert_\Min}$.
More technically, for a CPTP map $\Lambda$, let us write $\Pur(\Lambda)= 
 \{\rho_\Min\in \DensityOpSet{\Hilbert_\Min}\mid \tr(\Lambda[\rho_\Min]^2)=1 
 \}$ for the set of states which gets purified by the action of $\Lambda$.
For a purifying process of $\M$ we demand $\M \subset \Pur(\Lambda)$.
In this section we will only deal with the most simple non-trivial case where 
 only two states $\rho_1$ and $\rho_2$ are to be mapped onto a pure state, 
 i.e., $\M= \{\rho_1, \rho_2\}\subset \Pur(\Lambda)$.

Let us consider the case where we already have two purifying maps $\Lambda_A$ 
 and $\Lambda_B$ acting on $\rho_1$ and $\rho_2$, and without loss of 
 generality assume
\begin{equation}
 \td{\Lambda_A[\rho_1]}{\Lambda_A[\rho_2]}\ge 
 \td{\Lambda_B[\rho_1]}{\Lambda_B[\rho_2]}.
\end{equation}
(Here, $\td\rho\sigma=\tfrac12 \tr\abs{\rho- \sigma}$ with $\abs \LinearOp= 
 \sqrt{\LinearOp \LinearOp^\dag}$ denotes the trace distance of $\rho$ and 
 $\sigma$.)
Then there exists a CPTP map $\Omega$, such that up to a global unitary 
 transformation, $\Lambda_B[\rho_i]= (\Omega\circ \Lambda_A)[\rho_i]$ for 
 $i=1,2$:
For two pure states $\purestate{\psi_1}$ and $\purestate{\psi_2}$ one can 
 reduce the angle defined by $\sin\vartheta= 
 \td{\purestate{\psi_1}}{\purestate{\psi_2}}$ to an arbitrary angle $\varphi< 
 \vartheta$ via the CPTP map
\begin{subequations}\label{e1373}
\begin{equation}\label{e1373a}
 \Omega_{\varphi;\ket{\psi_1},\ket{\psi_2}}\colon \LinearOp\mapsto
  \sum_{\alpha =1}^3 A_\alpha\LinearOp A_\alpha^\dag,
\end{equation}
 with $A_\alpha$ being the Kraus operators \cite{Kraus:1983}
\begin{align}
 A_1&= \proj{\psi_1}+ a\proj{\psi^\perp_1},\label{e1373b}\\
 A_2&= (\sqrt{1-b^2} \ket{\psi_1}+
        \sqrt{b^2-a^2} \ket{\psi^\perp_1})\bra{\psi^\perp_1},\label{e1373c}\\
 A_3&= \id - \proj{\psi_1}- \proj{\psi_1^\perp},\label{e1373d}
\end{align}
\end{subequations}
 where ~ $a= \tan\varphi \cot\vartheta$, ~ $b= \sin\varphi/\sin\vartheta$, ~ 
  and $\ket{\psi^\perp_1}\in \Span\{\ket{\psi_1},\ket{\psi_2}\}$ is a 
  normalized vector orthogonal to $\ket{\psi_1}$.
Now ~ ~ let ~ $\purestate{\psi_i}= \Lambda_A[\rho_i]$ ~ and ~ choose ~ 
 $\sin\varphi= \td{\Lambda_B[\rho_1]}{\Lambda_B[\rho_2]}$.
Then, up to a global unitary transformation, $\Lambda_B[\rho_i]= 
 (\Omega_{\varphi;\ket{\psi_1},\ket{\psi_2}}\circ \Lambda_A)[\rho_i]$ holds.
Since we can mimic the action of $\Lambda_B$ on $\rho_1$ and $\rho_2$ by using 
 the map $\Lambda_A$, we would always prefer $\Lambda_A$ in favor of 
 $\Lambda_B$.
Thus among all purifying processes of two states we are most interested in 
 those which maximize the trace distance of the corresponding output states.

This trace distance of the output of a purifying map $\Lambda$ is upper bounded 
 by a geometric quantity depending on $\rho_1$ and $\rho_2$, namely by the 
 worst case distinguishability $\wcd{\rho_1}{\rho_2}$ \cite{Kleinmann:2006PRA},
\begin{equation}\label{e20664}
 \td{\Lambda[\rho_1]}{\Lambda[\rho_2]}\le \wcd{\rho_1}{\rho_2}.
\end{equation}
We now want to give a physical interpretation of this inequality:
In quantum mechanics, an ensemble of pure states $\purestate{\phi_j}$ with 
 probabilities $p_j> 0$ (where $\sum_j p_j=1$) is described by the mixed state 
 $\rho= \sum_j p_j \purestate{\phi_j}$.
In general, many different ensembles lead to the same density operator $\rho$, 
 and it is a prediction of quantum mechanics that it is impossible to 
 physically distinguish between such different ensembles.
Having said that, for a given mixed state $\rho$, a pure state $\purestate\phi$ 
 may physically appear if and only if $\purestate\phi$ can be part of an 
 ensemble that is represented by $\rho$, i.e., if and only if a positive number 
 $p$ exists, such that $\rho- p\purestate\phi$ is positive semidefinite.
Let us denote the collection of all such pure states $\purestate\phi$ by
\begin{equation}\begin{split}
 \mathcal Q_\rho
   &= \{ \purestate\phi \mid \exists p>0\colon \rho-p \purestate\phi \ge 0 \}\\
   &\equiv \{ \purestate\phi \mid \ket\phi\in \supp\rho \},
\end{split}\end{equation}
 where $\supp\rho$ is the support of $\rho$, i.e., the orthocomplement of the 
 kernel of $\rho$.
The worst-case distinguishability is now defined as
\begin{equation}\begin{split}
 \wcd{\rho_1}{\rho_2}&= \inf_{\purestate{\phi_i}\in \mathcal Q_{\rho_i}}
   \td{\purestate{\phi_1}}{\purestate{\phi_2}} \\
   &\equiv \min_k(\sin\vartheta_k),
\end{split}\end{equation}
 where $\vartheta_k$ denote the Jordan angles \cite{Stewart:1990} between 
 $\supp\rho_1$ and $\supp\rho_2$.

Let ~ us ~ continue ~ the ~ physical ~ motivation ~ of ~ Eq.~\eqref{e20664}.
The maximal success probability for distinguishing two mixed states via a 
measurement (``minimum error discrimination'') is given by 
\cite{Helstrom:1976,Herzog:2004PRA}
\begin{equation}
 P_\mathrm{MED}(\rho_1,\rho_2)= (1+\td{\rho_1}{\rho_2})/2,
\end{equation}
 where we assumed that both states have equal a priori probabilities.
Hence $P_\mathrm{MED}$ is the {\em average} success probability for 
 distinguishing the ensemble of pure states denoted by $\rho_1$ and $\rho_2$.
In a physical experiment, each single measurement is performed on a pure state 
 out of the ensembles, i.e., the task of the discrimination measurement is to 
 distinguish between a state in $\mathcal Q_{\rho_1}$ and a state in $\mathcal 
 Q_{\rho_2}$.
The optimal success probability to distinguish between such two pure states in 
 the {\em worst} case is given by
\begin{equation}\begin{split}
 P_\mathrm{WCD}&=
  \inf_{\purestate{\phi_i}\in \mathcal Q_{\rho_i}} 
         P_\mathrm{MED}(\purestate{\phi_1},\purestate{\phi_2})\\
  &\equiv (1+ \wcd{\rho_1}{\rho_2})/2.
\end{split}\end{equation}

Since no deterministic process can increase the trace distance between two 
 states \cite{Nielsen:2000}, a purifying process of $\rho_1$ and $\rho_2$ must 
 not deterministically increase the distance between any pair of pure states 
 $\purestate{\phi_1}\in \mathcal Q_{\rho_1}$ and $\purestate{\phi_2}\in 
 \mathcal Q_{\rho_2}$.
This may serve as a physical motivation for the inequality in 
 Eq.~\eqref{e20664}.

Can the bound in Eq.~\eqref{e20664} always be achieved by some purifying 
 process $\Lambda$?
The answer is affirmative, but in order to verify this to a satisfactory level 
 there is no way to avoid the awkwardness of an explicit construction of a map 
 which reaches equality in Eq.~\eqref{e20664}:

Let us first briefly recall the concept of Jordan bases and Jordan angles 
 (sometimes also called canonical bases and canonical angles) 
 \cite{Stewart:1990,Rudolph:2003PRA} of two subspaces $\mathcal A_1\subset 
 \Hilbert$ and $\mathcal A_2\subset \Hilbert$:
Orthonormal bases $\ket{\psi_1^k}$ of $\mathcal A_1$ and $\ket{\psi_2^k}$ of 
 $\mathcal A_2$ are called Jordan bases, if
\begin{subequations}
\begin{align}
 \bracket{\psi_1^k}{\psi_2^l}&= 0 &&\text{for $k\ne l$,}\\
 \bracket{\psi_1^k}{\psi_2^k}&=\cos\vartheta_k &&\text{for $k\le\min_i\dim 
  \mathcal A_i$}.
\end{align}
\end{subequations}
Such bases always exist and $\vartheta_k$ are called the Jordan angles between 
 $\mathcal A_1$ and $\mathcal A_2$.

The first step in the construction of the purifying map is to apply the 
 distance-decreasing map $\Omega_\varphi$ defined in 
 Eqns.~\eqref{e1373a}-\eqref{e1373d} on each pair of Jordan vectors 
 $\ket{\psi^k_1}\in \supp\rho_1$ and $\ket{\psi^k_2}\in \supp\rho_2$, such that 
 the ~ distance ~ is ~ reduced ~ to ~ $\wcd{\rho_1}{\rho_2}$:
We define the Kraus operators ~ $A_1^k$ ~ and ~ $A_2^k$ ~ for ~ $k\le 
 \min_i\rank \rho_i$ analogously to Eqns.~\eqref{e1373b} and \eqref{e1373c} and 
 choose $\sin\varphi_k= \wcd{\rho_1}{\rho_2}$.
In order to complete the set of Kraus operators, we in addition define the 
 projector $A_3= \id -\sum_{k} A_1^{k\dag} A_1^k- \sum_{k} A_2^{k\dag} A_2^k$ 
 and write
\begin{equation}\begin{split}
 \tilde\Omega&\colon
  \LinearOpSpace{\Hilbert_\Min}\rightarrow 
  \LinearOpSpace{\Hilbert_\Min}\\&\colon
  \LinearOp\mapsto \sum_{k} A_1^k\LinearOp A_1^{k\dag}+
    \sum_{k} A_2^k\LinearOp A_2^{k\dag}+A_3\LinearOp A_3^\dag,
\end{split}\end{equation}
Let $\purestate{\nu_i}$ be an arbitrary pure state in $\mathcal Q_{\rho_i}$.
One finds that
\begin{multline}
 \tilde\Omega[\purestate{\nu_i}]=
   \sum_k \tr(\purestate{\psi_i^k}\purestate{\nu_i})
            \tilde\Omega[\purestate{\psi_i^k}]\\+ A_3\purestate{\nu_i}A_3^\dag.
\end{multline}
By ~ construction, ~ $\purestate{\tilde\psi_i^k}= 
 \tilde\Omega[\purestate{\psi_i^k}]$ ~ is again ~ pure with 
 $\bracket{\tilde\psi_i^k}{\tilde\psi_j^l}= 0$ for $k\ne l$ and 
 $\td{\purestate{\tilde\psi_1^k}}{\purestate{\tilde\psi_2^k}}= 
 \wcd{\rho_1}{\rho_2}$.
Furthermore $A_3\purestate{\nu_i}A_3^\dag\ne 0$ only if 
 $\rank\rho_i>\rank\rho_j$ for $j\ne i$.

Using the above properties of $\tilde\Omega$, it is straightforward to find a 
 CPTP map $\tilde E\colon \Hilbert_\Min\rightarrow \Hilbert_\Min\otimes \Haux$, 
 such ~ that ~ the ~ vectors ~ $\ket k\ket{\phi_1}$ ~ diagonalize ~ $(\tilde 
 E\circ \tilde\Omega)[\rho_1]$ ~ and ~ the ~ vectors ~ $\ket k\ket{\phi_2}$ ~ 
 diagonalize ~ $(\tilde E\circ \tilde\Omega)[\rho_2]$, ~ where ~ $\bracket kl= 
 \delta_{kl}$ ~ and ~ $\td{\purestate{\phi_1}}{\purestate{\phi_2}}= 
 \wcd{\rho_1}{\rho_2}$.
Thus the map $\tr_\Min \circ \tilde E\circ \tilde\Omega$ is a map which reaches 
 the bound in Eq.~\eqref{e20664}, i.e.,
\begin{equation}\label{e20866}
 \wcd{\rho_1}{\rho_2}= \max_\Lambda \td{\Lambda[\rho_1]}{\Lambda[\rho_2]},
\end{equation}
 where the maximum is taken over all CPTP maps $\Lambda$ satisfying 
 $\{\rho_1,\rho_2\} \subset \Pur(\Lambda)$.
Furthermore, as already discussed in advance, due to Eq.~\eqref{e20866}, the 
 maximizing map $\tr_\Min\circ \tilde E\circ \tilde\Omega$ together with the 
 distance-decreasing map $\Omega_\varphi$ allows to mimic the action of any 
 purifying map of the states $\rho_1$ and $\rho_2$.

This result characterizes the output of any process, which maps two input 
 states to pure output states.
For example one immediately finds that two states with overlapping support have 
 a vanishing worst-case distinguishability and thus such states only can be 
 mapped to {\em identical} pure states by a purifying process.
In \cite{Kleinmann:2006PRA} the problem was investigated, how close the pure 
 output states of a purifying map can get to a purification 
 \cite{Hughston:1993PLA,Bassi:2003PLA} of the input states.
The deviation from the optimal quality of such a purifying map was found to be 
 limited by the difference $\td{\rho_1}{\rho_2}- \wcd{\rho_1}{\rho_2}$.
Furthermore the result in Eq.~\eqref{e20866} turned out to be the key for the 
 analysis of sets which can be mapped perfectly to their purifications 
 \cite{Kleinmann:2006PRA}.

In addition, the result in Eq.~\eqref{e20866} can also be used as a general 
 tool in quantum information theory, since results for pure states often are 
 much simpler to obtain than results for mixed states.
As an example, we provide a lower bound on the trace distance of any two 
 product states $\rho_1\otimes \sigma_1$ and $\rho_2\otimes \sigma_2$:
\begin{multline}\label{e16420}
 \td{\rho_1\otimes \sigma_1}{\rho_2\otimes \sigma_2}^2 \ge \\
    1-(1- \wcd{\rho_1}{\rho_2}^2)(1- \td{\sigma_1}{\sigma_2}^2).
\end{multline}
(From this inequality in particular $\td{\rho_1}{\rho_2}\ge 
 \wcd{\rho_1}{\rho_2}$ follows by setting $\sigma_1= \sigma_2$.)
This inequality follows by applying a map for which
\begin{align}
 \rho_1\otimes \sigma_1&\mapsto \purestate{\phi_1}\otimes
                                  (q_1\purestate 0+(1-q_1)\purestate 1),\\
 \rho_2\otimes \sigma_2&\mapsto \purestate{\phi_2}\otimes
                                  ((1-q_2)\purestate 0+q_2\purestate 1).
\end{align}
Such a mapping can be implemented by a CPTP map for appropriate $q_1$, $q_2$ 
 satisfying $q_1+q_2= 1+\td{\sigma_1}{\sigma_2}$ and $\purestate{\phi_i}$ 
 satisfying $\td{\purestate{\phi_1}}{\purestate{\phi_2}}= 
 \wcd{\rho_1}{\rho_2}$, since then for the first system one applies the 
 purifying map $\tr_\Min\circ \tilde E\circ \tilde\Omega$ and for the second 
 system one applies a minimum error discrimination of $\sigma_1$ and 
 $\sigma_2$.
Now using the fact that a CPTP map cannot increase the trace distance, it is 
 straightforward to obtain Eq.~\eqref{e16420}.

\section{Probabilistic Purifying Processes}\label{s18042}
Although we want to concentrate on deterministic processes, in this section we 
 wish to briefly discuss the properties of probabilistic purifying processes.
We exclude probabilistic processes $\bar\Lambda$ from our considerations, for 
 which $\tr\bar\Lambda[\rho]= 0$ for some $\rho\in \M$, i.e., we call a process 
 probabilistic on $\M$, only if for {\em any} state in $\M$ the success 
 probability of the process is non-zero.

A simple example of a probabilistic purifying process is a process, which first 
 performs an unambiguous state discrimination 
 \cite{Jaeger:1995PLA,Rudolph:2003PRA} between the possible input states and 
 then uses the unambiguous information to create a purification of the input 
 state:
In the language of probabilistic processes, unambiguous state discrimination of 
 a set of states $\rho_i$ is a probabilistic map which maps $\rho_i$ to $p_i 
 \purestate i$, where $p_i$ is the success probability of unambiguously 
 identifying $\rho_i$ and $\bracket ij= \delta_{ij}$.
In Ref.~\cite{Feng:2004PRA} it was shown, that a probabilistic unambiguous 
 state discrimination process for a set $\M$ exists, if and only if
\begin{equation}
 \supp\rho_i \nsubseteq \sum_{j\ne i} \supp\rho_j,
   \quad \forall \rho_i\in \M.
\end{equation}
Hence, if one applies unambiguous state discrimination on such a set $\M$, in 
 case of a successful discrimination one can map each state $\purestate i$ to a 
 purification $\purestate{\psi_i}$ of $\rho_i$.
This map is purifying as well as reversible on $\M$ (with the reversible map 
 being the partial trace over the purifying system) and it is successful, 
 whenever the unambiguous state discrimination process succeeds.

However, there is a good reason not to deepen the analysis of probabilistic 
 processes as a proper variant of purifying processes:
Physically, the information of a successful application of a probabilistic map 
 is provided as a bit of classical information.
Thus for a probabilistic purifying map $\bar\Lambda$ one can equivalently write 
 the deterministic map
\begin{equation}
 \Lambda\colon \LinearOp\mapsto \bar\Lambda[\LinearOp]+ (\tr\LinearOp 
 -\tr\bar\Lambda[\LinearOp]) \purestate?,
\end{equation}
 where $\purestate?$ is a state that is orthogonal to all output operators 
 $\bar\Lambda[\LinearOp]$.
However, the output of $\Lambda$ is not pure, unless $\bar\Lambda$ is already 
 {\em deterministic} and purifying.

\section{Purifying and Reversible Processes}\label{s29562}
We now want to combine the purifying property of a deterministic process with 
 the feature of reversibility.
Since we already noticed that processes which are reversible on all states 
 cannot increase the purity of any state (although it possible to decrease the 
 purity, e.g. using the map $E_\sigma$ defined in Eq.~\eqref{e18547}), in the 
 fashion of Section~\ref{s14898} we demand reversibility only on a subset of 
 states $\M$.
We call a CPTP map $\Lambda$ reversible on $\M$ if one can find a CPTP map 
 $\Lambda'$, such that $(\Lambda'\circ \Lambda)[\rho_\Min]= \rho_\Min$ for all 
 $\rho_\Min\in \M$.
Let us, again, formalize this property:
For a CPTP map $\Xi\colon \LinearOpSpace{\Hilbert_\Min}\rightarrow 
 \LinearOpSpace{\Hilbert_\Min}$, let $\Id(\Xi)= \{\rho_\Min\in 
 \DensityOpSet{\Hilbert_\Min}\mid \Xi[\rho_\Min]= \rho_\Min\}$ be the set of 
 states that are unchanged by the action of $\Xi$.
Thus for a reversible map $\Lambda$ on $\M$, we demand that one can find a CPTP 
 map $\Lambda'$, such that $\M\subset \Id(\Lambda'\circ \Lambda)$.
Note, that $\Lambda'$ does not need to be unique.
Now a map $\Lambda$ is purifying and reversible on $\M$, if and only if one can 
 find a map $\Lambda'$, such that $\M\subset \Pur(\Lambda)\cap 
 \Id(\Lambda'\circ \Lambda)$.
It is possible to completely characterize any such set $\M$:
\begin{theorem}\label{t11956}
A reversible and purifying process for a set of states $\M\subset 
 \DensityOpSet{\Hilbert_\Min}$ exists, if and only if for appropriate vector 
 spaces $\Hilbert_C^i$, $\Hilbert_A^i$ and $\Hilbert_B^i$, satisfying 
 $\Hilbert_\Min\otimes \Hilbert_C^i\cong \Hilbert_A^i\otimes \Hilbert_B^i$, one 
 can find mixed states $\sigma^i_B\in \DensityOpSet{\Hilbert_B^i}$ and 
 $\omega_C^i\in \DensityOpSet{\Hilbert_C^i}$, and unitary transformations 
 $U_i$, such that $\M= \bigcup_i \M_i$ with $\M_i\perp \M_j$, $i\ne j$ and
\begin{multline}\label{e27977}
 \{\rho\otimes \omega_C^i\mid \rho \in \mathcal M_i\}\subset\\
 \{U_i( \purestate\phi\otimes \sigma^i_B ) U_i^\dag \mid \ket\phi
   \in \Hilbert_A^i \}.
\end{multline}
\end{theorem}
In Theorem~\ref{t11956}, $\M_i\perp \M_j$ if $\tr(\rho\sigma)= 0$ for all 
 $\rho\in \M_i$ and $\sigma\in \M_j$.
Sets $\M_i$, which satisfy Eq.~\eqref{e27977} are called {\em essentially 
 pure}, i.e., a reversible and purifying process for $\M$ exists, if and only 
 if $\M$ is an orthogonal union of essentially pure sets.
Furthermore note, that basically, essentially pure sets are such sets which are 
 generated by applying the map $E_U\circ E_\sigma$ on a set of pure states.

\begin{proof}[Theorem~\ref{t11956}]
In \cite{Kleinmann:2006PRA} it was shown, that $\M$ is an orthogonal union of 
 essentially pure sets, if and only if a perfect purifier of $\M$ exists.
A perfect purifier is a CPTP map, which maps any state in $\M$ to one of its 
 purifications in $\DensityOpSet{\Hilbert_\Min\otimes \Haux}$.
Hence a perfect purifier $\Lambda$ of $\M$ in particular satisfies $\M\subset 
 \Pur(\Lambda)$ and $\M\subset \Id(\traux\circ \Lambda)$, i.e., it is purifying 
 and reversible.

For the converse assume that a reversible and purifying map $\Lambda$ for $\M$ 
 exists.
Then $\M\subset \Pur(\Lambda)$ and one can find a CPTP map $\Lambda'$, such 
 that $\M\subset \Id(\Lambda'\circ \Lambda)$.
The map $\Gamma_{\Lambda',\Lambda}$ defined in Eq.~\eqref{e25670} thus 
 satisfies $\Pur(\Gamma_{\Lambda',\Lambda})= \Pur(\Lambda)\supset \M$ and 
 $\Id(\traux\circ \Gamma_{\Lambda',\Lambda})= \Id(\Lambda'\circ \Lambda)\supset 
 \M$ and hence $\Gamma_{\Lambda',\Lambda}$ is a perfect purifier of $\M$.
Using again the result in \cite{Kleinmann:2006PRA}, it follows that $\M$ is an 
 orthogonal union of essentially pure sets.\qed
\end{proof}

Although Theorem~\ref{t11956} completely characterizes all sets for which a 
 reversible and purifying process exists, it is in general not straightforward 
 to test whether a set is of the structure as specified in Eq.~\eqref{e27977}.
Only for the case where $\M$ consists of only two states, an operational 
 necessary and sufficient criterion is known \cite{Kleinmann:2006PRA}:
The set $\M=\{\rho_1, \rho_2\}$ is essentially pure or $\rho_1\perp \rho_2$ if 
 and only if $\wcd{\rho_1}{\rho_2}= \td{\rho_1}{\rho_2}$.
In the general case only some necessary operational conditions can be derived.
The most obvious necessary criterion is, that in an essentially pure set all 
 states must share the same spectrum.
Another example of a necessary criterion is, that the Jordan angles between the 
 support of any two states taken from an essentially pure set have to be 
 completely degenerate.
But these two properties are not sufficient for an essentially pure set, as the 
 following simple counter-example demonstrates:
\begin{subequations}
\begin{align}
 \rho_1&= p\purestate0+ (1-p)\purestate1 \\
 \rho_2&= p\purestate{\nu^+}+ (1-p) \purestate{\nu^-},
\end{align}
\end{subequations}
 where $\ket{\nu^\pm}= \frac12 (\pm\ket0+ \ket1\pm \ket2+ \ket3)$ and 
 $0<p<\frac12$.

As a final remark let us note that it is possible to simplify the definition of 
 essentially pure sets:
A set of states $\M \subset \DensityOpSet{\Hilbert_\Min}$ with $\rho_0\in \M$ 
 is essentially pure if and only if one can find a unitary transformation $U$ 
 on $\Hilbert_\Min\otimes \Haux$ and normalized vectors $\ket\rho\in \Haux$ 
 corresponding to each $\rho\in \M$, such that for each $\rho\in \M$,
\begin{equation}\label{e25064}
 \rho \otimes \purestate{\rho_0}= U( \rho_0\otimes \purestate\rho) U^\dag
\end{equation}
 holds.
From the proof of Theorem~1 in \cite{Kleinmann:2006PRA} it is clear that in 
 Eq.~\eqref{e27977} one always can choose $\omega_C$ to be pure.
Now the dimension of the kernel of each element on the left hand side of 
 Eq.~\eqref{e27977} is given by $\dim(\Hilbert_\Min)\dim(\Hilbert_C)-\rank\rho$ 
 while on the right hand side we find 
 $\dim(\Hilbert_A)\dim(\Hilbert_B)-\rank\rho$.
One readily extends $\Hilbert_C$ and $\Hilbert_B$ such that $\dim\Hilbert_B$ is 
 an integer multiple of $\dim(\Hilbert_\Min)$.
Then after a suitable rotation $U'$ on $\Hilbert_A\otimes \Hilbert_B$, one has 
 $U'(\purestate\phi\otimes \sigma_B ){U'}^\dag= (\rho_0\otimes \purestate0) 
 \otimes \purestate\phi$.
Identifying $\purestate0\otimes \purestate\phi$ with $\purestate\rho$ finishes 
 the proof of Eq.~\eqref{e25064}.

\section{Conclusions}\label{s10522}
In summary we have analyzed deterministic physical processes which are
 reversible or purifying, with particular focus on the combination of both 
 properties.
First we have shown that the properties of reversibility and purity of a 
 physical processes are completely incompatible, as long as reversibility or 
 purity is required to hold for any input state.
For certain restricted sets, however, one can combine these properties.
We investigated the case, where only two input states are mapped to pure 
 states.
It turned out that the trace distance of the output states of such a map is 
 limited by the worst-case distinguishability of the input states.
A map was provided, which always reaches this bound.
Some applications of this result in quantum information theory were presented.
For {\em probabilistic} processes we used unambiguous state discrimination to 
 build a non-trivial example of a purifying and reversible process.
We finally characterized all sets of states, for which a {\em deterministic} 
 purifying and reversible process exists and it turned out that such sets have 
 to be pure up to a common mixed contribution.
Despite this result and the existence of an operational criterion for such 
 essentially pure sets in the case, where the set consists of only two states, 
 no operational necessary and sufficient condition for larger essentially pure 
 sets was provided.
Such criteria will be subject to further research.
Furthermore, although some properties of reversible processes where presented 
 here, another direction of future work will be to deepen the understanding of 
 such processes.

\begin{acknowledgement}
This work was partially supported by the European Commission (Integrated 
 Projects SECOQC and SCALA).
\end{acknowledgement}

\bibliography{the}

\begin{thebibliography}{10}

\bibitem{Stinespring:1955PAMS}
W.~Forrest Stinespring.
\newblock Positive functions on ${C}^*$-algebras.
\newblock {\em Proc. Amer. Math. Soc.}, 6:211--216, 1955.

\bibitem{Nielsen:2000}
Michael~A. Nielsen and Isaac~L. Chuang.
\newblock {\em Quantum Computation and Quantum Information}.
\newblock Cambridge University Press, Cambridge, 2000.

\bibitem{Kraus:1983}
K.~Kraus.
\newblock {\em States, Effects and Operations}.
\newblock Springer-Verlag Berlin, 1983.

\bibitem{Kleinmann:2006PRA}
Matthias Kleinmann, Hermann Kampermann, Tim Meyer, and Dagmar Bru\ss.
\newblock Physical purification of quantum states.
\newblock {\em Phys. Rev. A}, 73:062309, 2006.

\bibitem{Stewart:1990}
Gilbert~W. Stewart and Ji-guang Sun.
\newblock {\em Matrix Pertubation Theory}.
\newblock Academic Press, San Diego, 1990.

\bibitem{Helstrom:1976}
Carl~W. Helstr{\o}m.
\newblock {\em Quantum Detection and Estimation Theory}.
\newblock Academic, New York, 1976.

\bibitem{Herzog:2004PRA}
Ulrike Herzog and J\'{a}nos~A. Bergou.
\newblock Distinguishing mixed quantum states: Minimum-error discrimination
  versus optimum unambiguous discrimination.
\newblock {\em Phys. Rev. A}, 70:022302, 2004.

\bibitem{Rudolph:2003PRA}
Terry Rudolph, Robert~W. Spekkens, and Peter~S. Turner.
\newblock Unambiguous discrimination of mixed states.
\newblock {\em Phys. Rev. A}, 68:010301(R), 2003.

\bibitem{Hughston:1993PLA}
Lane~P. Hughston, Richard Jozsa, and William~K. Wootters.
\newblock A complete classification of quantum ensembles having a given density
  matrix.
\newblock {\em Phys. Lett. A}, 183:14--18, 1993.

\bibitem{Bassi:2003PLA}
Angelo Bassi and GianCarlo Ghirardi.
\newblock A general scheme for ensemble purification.
\newblock {\em Phys. Lett. A}, 309:24--28, 2003.

\bibitem{Jaeger:1995PLA}
Gregg Jaeger and Abner Shimony.
\newblock Optimal distinction between two non-orthogonal quantum states.
\newblock {\em Phys. Lett. A}, 197:83--87, 1995.

\bibitem{Feng:2004PRA}
Yuan Feng, Runyao Duan, and Ying Mingsheng.
\newblock Unambiguous discrimination between mixed quantum states.
\newblock {\em Phys. Rev. A}, 70:012308, 2004.

\end{thebibliography}
\end{document}